\documentclass[aps,preprint,groupedaddress,showpacs]{revtex4}
\usepackage{graphicx}

\newcommand{\e}{\epsilon}
\newcommand{\ep}{\epsilon^\prime}

\begin{document}
\bibliographystyle{apsrev}

\preprint{}

\title{High-Energy Neutrinos from Photomeson Processes in Blazars}

\author{Armen Atoyan$^1$}\author{Charles D. Dermer$^2$}
\affiliation{$^1$CRM, Universite de Montreal, Montreal H3C 3J7, Canada \\
$^2$Code 7653, Naval Research Laboratory, Washington, DC 20375-5352}

\date{\today}

\begin{abstract}

An important radiation field for photomeson neutrino production in 
blazars is shown to be the radiation field external to the jet. 
Assuming that protons are accelerated with the same power as electrons and 
injected with a $-2$ number spectrum, we predict 
that km$^2$ neutrino telescopes will detect $\gtrsim $ 1 neutrinos per year 
from flat spectrum radio quasars (FSRQs) such as 3C 279. The escaping 
high-energy neutron and photon beams transport inner jet energy far from the 
black-hole engine, and could power synchrotron X-ray jets and FR II hot spots 
and lobes.
\end{abstract}

\maketitle

EGRET observations of $\approx 100$ MeV - 5 GeV emission from over 60 FSRQs 
and BL Lac objects established that blazars are, with gamma-ray bursts,
among the most powerful accelerators of relativistic particles in 
nature \cite{har99}.  The standard blazar model consists of a supermassive
black hole ejecting twin jets of relativistic plasma, one of which is
pointed towards us. 
Observations of bright, highly variable 
nonthermal radiation 
from blazars mean that regions of intense 
photon and nonthermal particle energy densities, both of which are needed for
efficient photopion production \cite{mb92,mpr01}, are found in these sources. 
Blazars therefore represent a potential source of energetic $\pi^{\pm}$-decay  
neutrinos 
to be detected by operating and planned high-energy neutrino telescopes
\cite{ghs95}.  Previous treatments have considered 
internal synchrotron photons and the direct disk radiation field \cite{man93}  
as targets for high-energy proton interactions in jets, 
and neutrino production in the cores of AGN (active galactic nuclei) \cite{ste91}. Here we show that photons
from external quasi-isotropic radiation fields, which have earlier been proposed as
target photons to be Compton-scattered by nonthermal electrons to
$\gamma$-ray energies \cite{sbr94}, also provide the most important
photon source for photomeson production of $\gtrsim 30$ TeV neutrinos
in FSRQs. The inclusion of this effect increases neutrino detection
rates by more than an order-of-magnitude.  Moreover, the neutrinos are
formed at energies $\gtrsim 3\times 10^{13}$ eV rather than at
$\gtrsim 10^{17}$ eV \cite{mp01}, which improves prospects for
detection.  

Strong optical emission lines \cite{net90} from the
illumination of broad-line region (BLR) clouds reveal 
bright accretion-disk and scattered disk radiation \cite{sbr94} in the inner regions of FSRQs. 
BL Lac objects have weak emission lines, so the dominant soft photon source is thought to 
be the internal synchrotron emission in BL Lac objects. 
In our analysis of photomeson production in FSRQs, we assume that
the quasi-isotropic scattered external
radiation field dominates the direct accretion-disk field.
The photomeson neutrino spectrum can be calculated once the
mean magnetic field and comoving spectral energy density are determined. The
measured variability time scale $t_{var}$ and synchrotron ($L_s$) and 
Compton ($L_C$) luminosities determined from the spectral energy distribution of the
bright, well-studied blazar 3C 279  \cite{har96} implies physical parameters
of its jet emission region. Its redshift $z = 0.538$, which
implies a luminosity distance $d_{\rm L}\cong 1.05\times10^{28}\,\rm
cm$ for an $\Omega_m = 0.3$, $\Omega_\Lambda = 0.7$ cosmology with a
Hubble constant of 65 km s$^{-1}$ Mpc$^{-1}$.  A crucial unknown is
the Doppler factor $\delta = [\Gamma(1-\beta_\Gamma\cos\theta)]^{-1}$,
where $\Gamma$ is the bulk Lorentz factor of the relativistic plasma
blob, and $\theta$ is the angle between the jet and observer
directions. 

The comoving synchrotron photon energy density is given by $u^\prime_s
\cong L_s/(2\pi r_b^2 c \delta^4)$, where $r_b$ is the comoving radius of the
blob, here assumed spherical, and primes denote comoving
quantities. The spectral energy density 
$u_s(\epsilon^\prime)\equiv
m_e c^2\epsilon^\prime n^\prime_s(\epsilon^\prime)$ 
(units of ergs s$^{-1}\epsilon^{\prime - 1 }$) of photons with
dimensionless comoving energy $\ep = h\nu^\prime/m_ec^2$ is found through
\begin{equation}
 \epsilon^\prime u_s^{\prime}(\epsilon^\prime)\; \cong 
\frac{2 d_{\rm L}^2  f_s(\epsilon)}
{r_{\rm b}^2 \, c \, \delta^4}\;\cong 
\frac{2 d_{\rm L}^2  (1+z)^2 f_s(\epsilon)}
{ c^3 \, t_{var}^2\delta^6}\; \;,
\label{eq1}
\end{equation}
in terms of the $\nu F_\nu$ flux $f_s(\epsilon)$ of the synchrotron
component. Here $\epsilon = \delta\epsilon^\prime /(1+z)$, and 
we relate $t_{var}$ and $r_b$ through the expression $r_b
\cong ct_{var}\delta/(1+z)$.

For calculations of $n^\prime_s(\ep )$ from 3C 279, we approximate the
flux density $F(\e)\propto \e^{-\alpha}$ observed during the flare of 1996
\cite{weh98} in the form of a continuous broken power-law function,
with indices $\alpha_{1}\cong 0.5$, $\alpha_{2}\cong
1.45$, and $\alpha_{3}\cong 0.6$ at frequencies $\nu_0 < \nu \leq
\nu_1=\nu_{pk}$, with $\nu_{pk} \cong 10^{13}\nu_{13}\,\rm Hz$, 
$\nu_1 \leq \nu
\leq \nu_2 =10^{16}\,\rm Hz$, and $\nu \geq \nu_2$, respectively.  The
$\nu F_\nu$ synchrotron radiation flux $f_s(\epsilon)$ reaches a maximum 
value $f_s(\epsilon_{pk}) \cong 1.7\times 10^{-10} \,\rm erg\, cm^{-2}\,
s^{-1}$ at $\nu = \nu_{pk}$ or $\e = \e_{pk}$. The flaring and 
three-week average flux of the
Compton $\gamma$-ray component peaks at $\sim 500 \,\rm MeV$ and
 is $\sim 20$ and $\sim 5$-10 times larger than 
$f_s(\epsilon_{pk})$ at the corresponding times \cite{weh98}.
Using $t_{var}(d)\approx 1 $ (in days) for 3C 279
as observed by EGRET \cite{weh98} during the three week average, 
one finds from equation (\ref{eq1}) with 
$\delta_{10}\equiv \delta /10$ that
\begin{equation}
u_s^{\prime}
\simeq \e_{pk}^\prime u_s^\prime(\e^\prime_{pk}) 
\simeq 0.4 \, [t_{\rm var}({\rm d})]^{-2} \, \delta_{10}^{-6}
\; {\rm erg~cm}^{-3} \, .
\end{equation} 

If the $\gamma$-rays from FSRQs are due to Compton-scattered radiation
from external photon fields, then $L_{EC}/L_s \cong
u^\prime_{ext}/u_B$ \cite{sbr94}, where $L_{EC} = L_C - L_{SSC}$
is the measured power from Compton-scattered external photon fields, and 
$L_{SSC}$ is the synchrotron-self-Compton (SSC) power.
Consequently $u^\prime_{ext} \cong u_B (L_{EC}/L_s) = a u^\prime_s = a
L_s/(2\pi r_b^2c
\delta^4)$, where $a \equiv L_{EC}/L_{SSC}$. The energy of the external
photons in the comoving frame is $\e_{ext}^\prime \cong
\delta\e_{ext}$. Models taking into account SSC and external Compton
(EC) components show that a complete spectral fit requires
synchrotron, SSC and EC components \cite{bot99}, with $a \sim 0.1$-1
for BL Lac objects and $a \sim 1$-10 for FSRQs. During the 1996 flare
of 3C 279, $a \simeq 10$, so that the energy density of external soft
photons in the comoving blob frame is $\sim 10\times$ greater
than the synchrotron energy density. Moreover, the energy of these
photons increases with $\delta$, lowering the photomeson production threshold.

Lower limits to $\delta$ are defined by the condition that the
emitting region be transparent to $\gamma\gamma$ pair-production
attenuation inside the blob 
\cite{ghi92}, which requires that
$\tau_{\gamma\gamma}(\epsilon^\prime) \cong 2\sigma_{\rm
T}n^\prime_s(2/\epsilon^\prime)r_b/(3\epsilon^\prime) < 1$.
 The measured $> 10^{16}$ Hz flux during the 1996 flare then 
implies that $\delta \geq 5 [E_{ph}({\rm
GeV})^{0.12}/[t_{var}({\rm d})]^{0.19}$, 
where $E_{ph}$(GeV) is the photon energy in GeV. 

The magnetic field $B$ in the blob can be determined by introducing the
equipartition parameter $\eta = u^\prime_{el}(1+k_{pe})/u_{\rm B}$ for the
ratio of relativistic
electron to magnetic energy densities in the jet, with the factor
$k_{pe}$ correcting for the presence of nonthermal hadrons.
We assume $k_{pe}$ = 1 in the estimate of $B$.
The measured synchrotron flux density $F_s \propto
\e^{-\alpha}$ with $\alpha\simeq 0.5$ 
at $ \nu \lesssim 10^{13}\,\rm Hz$ gives the equipartition magnetic field
\begin{equation}
B({\rm Gauss})\cong 130 \;{d_{28}^{4/7}f_{-10}^{2/7}
[(1+k_{pe})\ln(\nu_0/\nu_1)]^{2/7}(1+z)^{5/7}\over \eta^{2/7}
[t_{\rm var}({\rm d})]^{6/7}\delta^{13/7}\nu_{13}^{1/7} }\;
\label{Beq}
\end{equation}
where $\nu_0/\nu_1 \cong 10^3$ and $f_{-10} \equiv$ 
$ f_s(\e_{pk})/(10^{-10}$erg cm$^{-2}$
s$^{-1})= 1.7$ for 3C 279.

An alternative estimate of $B$ can be derived from the observed synchrotron
and Compton powers \citep{tav98}. Neglecting Klein-Nishina effects on
 Compton scattering, the ratio of the Compton and synchrotron powers
 $L_C/L_s \cong u^\prime_{ph}/u_B$, so that 
$u_B = (L_s/L_C)u^\prime_s(1+ u^\prime_{ext}/u^\prime_s) = 
(L_s/L_C)u^\prime_s(1+a)$.
 Using equation (\ref{eq1}) in this expression for $B$ gives
\begin{equation}
B({\rm Gauss}) \cong {2(1+z)\over \delta^3 t_{var}}\;{L_s \sqrt{1+a}\over c^{3/2}L_C^{1/2}}\;.
\label{B2}
\end{equation} 

The observed flux of synchrotron radiation from 3C 279 with 
$\alpha = 0.5 $ at $\nu \ll 10^{13} \,\rm Hz$ implies that 
$ B(\rm G) \simeq 7 \;\eta^{-2/7} [t_{\rm var}(\rm d)]^{-6/7} \, 
\delta_{10}^{-13/7}$
from equipartition arguments. The estimate of $B$ from equation
(\ref{B2}) gives $B(\rm G) \cong 4\; t^{-1}_{var}(\rm
d)\delta_{10}^{-3}$, using $L_s = 3\times 10^{47}$ ergs s$^{-1}$,
$L_C/L_s \cong 10$, and $a \cong 10$.
The external radiation energy density in the comoving frame, using $B$ 
derived from equation (\ref{Beq}), is
$u_{ext}^{\prime}\simeq 1.5 \, (L_{EC}/ L_s)
 \eta^{-4/7} 
[t_{\rm var}(\rm d)]^{-12/7}\delta_{10}^{-26/7} \,{\rm erg~cm}^{-3}$.
Note the weaker dependence of $u^\prime_{ext}$ on $\delta$ compared to 
$u^\prime_s$. For the calculations we  take $\eta=1$ and   
$L_{EC}/ L_s = 10$.

Energy losses of relativistic protons (and neutrons) are calculated on
the basis of standard expressions (e.g., Ref.\ \cite{bg88}) for the
cooling time of relativistic protons due to photopion production in $p
\gamma$ collisions. If the ambient photons have spectral density
$n^\prime_{\rm ph}(\ep) $, then $$ t_{\rm p \gamma}^{-1}(\gamma_p) =
\int_{\frac{\epsilon_{\rm th}} {2 \gamma_p}}^{\infty}{\rm d}\ep\;{
\frac{c \, n^\prime_{\rm ph}(\ep)}{2 \gamma_p^2 \epsilon^{\prime 2} }
\int_{\epsilon_{\rm th}}^{2\ep \gamma_p}{\rm d} \epsilon_{\rm r}\;{
\sigma(\epsilon_{\rm r}) K_{\rm p \gamma}(\epsilon_{\rm r}) \epsilon_{\rm r} 
  }\, }
\, , $$ 
where $\gamma_p$ is the proton Lorentz factor, $\epsilon_{\rm r}$
is the photon energy in the proton rest frame, $\sigma(\epsilon_{\rm
r})$ is the photopion production cross section, $\epsilon_{\rm
th}\approx 150 \,\rm MeV$ is the threshold energy for the parent
photon in the proton rest frame, and $ K_{\rm p \gamma}(\epsilon_{\rm
r})$ is the inelasticity of the interaction.  The latter increases
from $ K_{1}\approx 0.2 $ at energies not very far above the threshold, 
to $K_{2}\sim
0.5-0.6 $ at larger values of $\epsilon_{\rm r}$ where multi-pion
production dominates \cite{bg88,muc99}.

A detailed recent study of this photohadronic process is given
by Ref.\ \cite{muc99}. To simplify calculations, we
approximate $\sigma(\epsilon_{\rm r})$ as a sum of 2
step-functions $\sigma_{1}(\epsilon_{\rm r})$ and
$\sigma_{2}(\epsilon_{\rm r})$ for the total single-pion 
($p + \gamma \rightarrow p+\pi^0$ or $n + \pi^+$) 
and multi-pion channels, respectively, with $\sigma_{1}=380~ \mu \rm b$
for $ 200 \,\rm MeV \leq \epsilon_{\rm r}
\leq 500 \, MeV$ and $\sigma_1 = 0$ outside this region, whereas
$ \sigma_2 = 120 ~\mu \rm b$ at $\epsilon_{\rm r} \geq 500 \,\rm MeV$.
The inelasticity is approximated as $K_{\rm p\gamma} =K_1$ and $K_{\rm
p\gamma}=K_2$ below and above 500 MeV.  Our
calculations give good
agreement with more detailed treatments of
the time scales for photopion interactions of ultra-relativistic cosmic
rays with the cosmic microwave background \cite{bg88,sta00}. 
This approach also works well for a broad
power-law distribution of field photons $u^\prime_{\rm ph}(\ep )
\propto (\epsilon^{\prime})^{-\alpha}$ for different  spectral 
indices $\alpha$, and readily explains the significant
increase in the mean inelasticity of incident protons (or neutrons)
from $\langle K_{\rm p\gamma}\rangle \simeq 0.2 $ for steep photon
spectra with $ \alpha_{\gamma} \gtrsim 1 $, to
$\langle K_{\rm p\gamma}\rangle \rightarrow 0.6 $ for hard spectra
with $ \alpha_{\gamma} < 1 $ \cite{muc99}.

The spectra of secondary $\pi^{0,\pm}$-decay particles ($\nu,
\; \gamma, \; e$) are calculated in the $\delta$-function approximation,
assuming that the probabilities for producing pions of different
charges ($\pi^0$, $\pi^{+}$ and $\pi^{-}$) are equal for the
multi-pion interaction channel. To correctly apply the
$\delta$-function approximation, one has to properly take into account
the different inelasticities of the multi-pion and single-pion
production channels (see Ref.\ \cite{muc99} for comparison).

We assume that the spectrum of the external UV radiation field arises
from a Shakura/Sunyaev \cite{ss73} optically-thick accretion disk model
that is scattered by BLR clouds, though other sources of external photons
could be reprocessed line emission or infrared radiation from a torus. 
This disk model has flux density $F(\epsilon ) \propto
\epsilon^{1/3}\exp(-\e/\e_{max})$, where the maximum photon energy $\e_{max}$ 
is determined by the innermost radius of the blackbody disk and
properties of the central engine. We take $m_ec^2 \e_{max} = 35$ eV
\cite{ds93}.  

Fig.\ 1 shows observer-frame energy loss timescales  of
protons due to photopion production in a jet of 3C~279 calculated for
Doppler-factors $\delta = 7$, 10, and 15. 
For comparison, we show also the 
photopion timescales if only the interactions with the synchrotron radiation
were taken into account for the cases $\delta =7$ and 10. 
Fig.\ 2 shows the energy distributions of relativistic protons $N_{\rm
p}(E)$ in the jet plasma blobs 
calculated assuming power-law injection of 
relativistic protons with
number index $\alpha_{\rm p}=2$ on observed timescales $\Delta t = 3
\,$ weeks. The total injection power of the protons $L_{p}=
2\times 10^{48}\delta^{-4}$ erg~s$^{-1}$, which corresponds to the average
measured $\gamma$-ray luminosity from 3C 379 during the three week observing
period surrounding the 1996 flare \cite{weh98}.  In calculations of $N_{\rm
p}$ we take into account the photohadron interaction energy losses, as
well as the escape losses of the protons in the Bohm diffusion limit.
This limit ensures that the gyroradii
$r_{\rm L} (\rm cm) \cong 3.1\times 10^6 (A \gamma_p /Z)/B({\rm G})$
of the highest energy ions do not exceed $r_b$ \citep{hil84}.

Importantly, we also take into account proton losses through
the $p\gamma \rightarrow nX$ channel.
The neutrons may then escape the blob either before they decay or collide
again with photons inside the blob.
To demonstrate the significance of this channel, in Fig.\ 2 
by the full dotted curve
we show the proton distribution calculated for 
the case of $\delta =7$, but neglecting the effect 
of escaping neutrons. Comparison with the solid 
curve indicates that $\simeq 50$\% of the $E\geq 10^{14}
\, \rm eV$ proton energy 
is carried by relativistic neutrons from the 
blob, forming a highly collimated neutron beam with opening angle 
$\theta_{b} \sim 1/\Gamma$ 
and a total power $L_{n}^{(beam)} \sim 0.25 L_{p} 
\sim 5\times 10^{47}\delta^{-4}\,\rm erg\, s^{-1}$ for the parameters 
assumed in Fig.~2. 
  
Fig.\ 3 shows the expected differential fluences
of neutrinos produced by protons in Fig.\ 2,  integrated 
over the 3 week flaring period. Note that the
mean neutrino energy fluxes at $E\geq 100\,\rm TeV$ are at the 
level  $E \, F_{\nu}(E) \simeq E^2 \Phi_\nu /\Delta t \lesssim 2 
\times 10^{-11} \rm \, erg \, cm^{-2} \, s^{-1}$. 
Given very similar emissivities in overall secondary $\nu$ 
and $(\gamma + e)$ production (e.g. Ref. \cite{muc99}), the {\it unabsorbed}
fluxes of multi-TeV $\gamma$-rays would therefore be at the same level,
which makes only $\leq 2\%$ of the $\gamma$-ray energy flux 
detected by EGRET. A pair-photon cascade in the blob would increase this 
fraction to the level of $\leq 10\%$ contribution of the cascade 
radiation to the detected $\gamma$-ray flux, reprocessing most of 
the initial $\gamma$-ray energy into $\gamma$-radiation with  
$E\lesssim 10\,\rm GeV$.
A small fraction of the energy will appear as  
synchrotron radiation in the X-ray to GeV $\gamma$-ray band. 
Because the cascade synchrotron power is emitted with a very hard
spectrum over a wide energy range, it will not
overproduce also the observed X-ray flux of 3C~279.
These estimates show, however, that the nonthermal proton power 
cannot be more than $10 \times$ the $\gamma$-ray power, 
or the cascade radiation would exceed the observed fluxes.

For the fluences in Fig.4,
 the total number $N_\nu$ of
neutrinos that could be detected by a $1 \,\rm km^2$ detector, 
using neutrino detection efficiencies given by Ref.\ 
\cite{ghs95}, is 0.45, 0.27 and 0.12 for the cases of $\delta=7$, 10
and 15, respectively. 
These numbers are, however, much less if the external field
is neglected,
giving instead the values $N_\nu \approx 0.055$ ($\delta =7$)
and $0.014$ ($\delta = 10$) for the
fluences in Fig.\ 3. This would not leave a realistic prospect for the
detection of at least 2-3 neutrinos, which is required for a positive
detection of a source. BL Lac objects, which have weak BLRs and, therefore, 
a weak scattered external radiation field,
should consequently be much weaker neutrino sources.

For a 10\% flaring duty factor, and
considering the additional neutrino production during the quiescent
phase, we expect that IceCube or other km$^2$ array may
detect several neutrinos per year from 3C 279-type blazar jets with
$\delta \approx 5$-15.  Allowing an overall power of protons
larger than that for primary electrons improves the prospects for neutrino 
detection. A softer nonthermal proton spectrum would, however, reduce
the 
neutrino emissivity.

Acceleration of protons only to
$E_{\rm p}\gtrsim 10^{14}\,\rm eV$ is required for efficient neutrino
production through photomeson interactions, as can be seen from Fig.1.
Thus we predict that a km$^2$ array will detect
high-energy neutrinos from FSRQs without requiring acceleration of
protons to ultra-high ($\gtrsim 10^{19}$ eV) energies.
The maximum cosmic-ray energies from blazars are found to be 
$E_{CR,max} \simeq 8\times 10^{17} Z\Gamma^2 B({\rm G}) 
t_{var}({\rm d})/(1+z)$ eV \cite{hil84}, 
with maximum neutrino energies a factor $\sim 20$ lower. Given 
additionally the intense 
photomeson interactions encountered by a proton 
at this optimistic limit in the jet environment 
(see Fig.\ 1), it is unlikely that the inner jets of blazars are
 significant sources of ultra-high energy  cosmic rays.  
The neutrino production from FSRQs is therefore unaffected by the 
cosmic ray bound of Ref.\ \cite{wb98} (though note the more general 
upper limit derived in Ref.\ \cite{mpr01}).

About half of the photomeson interactions in the plasma jet will produce 
neutrons. Neutrons with comoving 
Lorentz factors $\gamma_n^\prime\gtrsim 10^3 \delta_{10} t_{var}({\rm d})$ 
will escape the blob in a collimated beam, and continue 
to interact efficiently with the external UV radiation field up to 
$\sim 0.1-1$ pc scales, initiating pair-photon cascades far away 
from, but in the same direction with, the inner jet due to secondary 
$\gamma$-rays and electrons. 
FSRQ jets will thus form intense neutron and photon beams that transport 
energy far from the central engine of the AGN \cite{ato92}.
Cosmic-ray neutrons with energies of $10^{17}E_{17}\,$eV will decay
at distances $\lesssim E_{17}$ kpc from the jet core. 
The deposition of the energy of the beamed
hadrons and multi-TeV $\gamma$-rays (through $\gamma
\gamma$-absorption at $>10\,\rm kpc$ scales) 
can reasonably be at the mean rate 
$\sim 10^{43} \,\rm erg\, s^{-1}$, which could lead to enhanced 
X-ray synchrotron emission along the large-scale jet,
as seen by the {\it Chandra X-ray Observatory} from sources such as 
Pictor A \cite{wys01}. Depending on the stability of charged particle 
transport, nonthermal particles could be deposited at $\gtrsim 100$ kpc 
to supply power for hot spots and lobes in
FR II galaxies, though {\it in situ} particle acceleration could also take
place at these sites. The predicted weaker 
neutron and neutrino production in BL Lac objects and their parent FR I 
galaxies might account for the morphological differences
of these radio galaxy types. 
This picture will be tested by sensitive high-energy neutrino detectors, 
ground-based $\gtrsim 50$ GeV gamma-ray detectors, 
and {\it GLAST} \cite{ghs95,gla01}.

\vskip0.2in
\noindent AA appreciates the support and hospitality 
of the NRL Gamma and Cosmic Ray Astrophysics Branch during his visit
 when this work was initiated.
The work of CD is supported by the Office of Naval Research and NASA 
grant DPR S-13756G. 

\bibliography{your bib file}

\begin{figure}
{\includegraphics{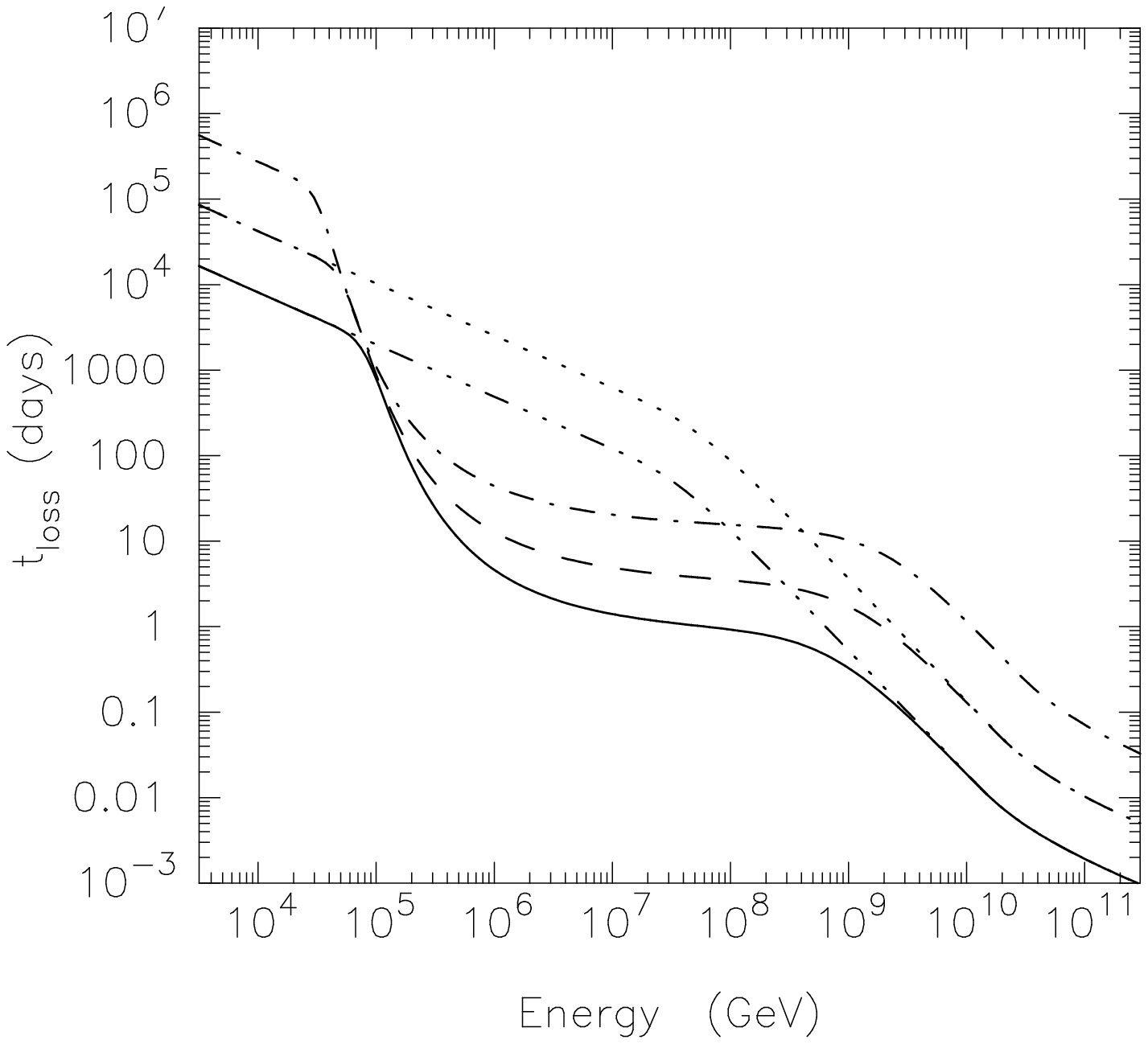}}
\caption[]{
Photomeson interaction energy loss time scale of protons, calculated
for spectral fluxes observed from 3C~279 (see text) and $t_{\rm
var}=1\,\rm d$, assuming 3 different Doppler-factors for the jet:
$\delta = 7$ (solid curve), $\delta = 10$ (dashed curve), and $\delta
= 15$ (dot-dashed curve). The dotted and triple-dot--dashed curves are
calculated for $\delta = 7$ and $\delta = 10$, respectively, when $p
\gamma$ interactions with the synchrotron radiation field alone are
considered.}
 \label{Fig1}
\end{figure}

\begin{figure}
{\includegraphics{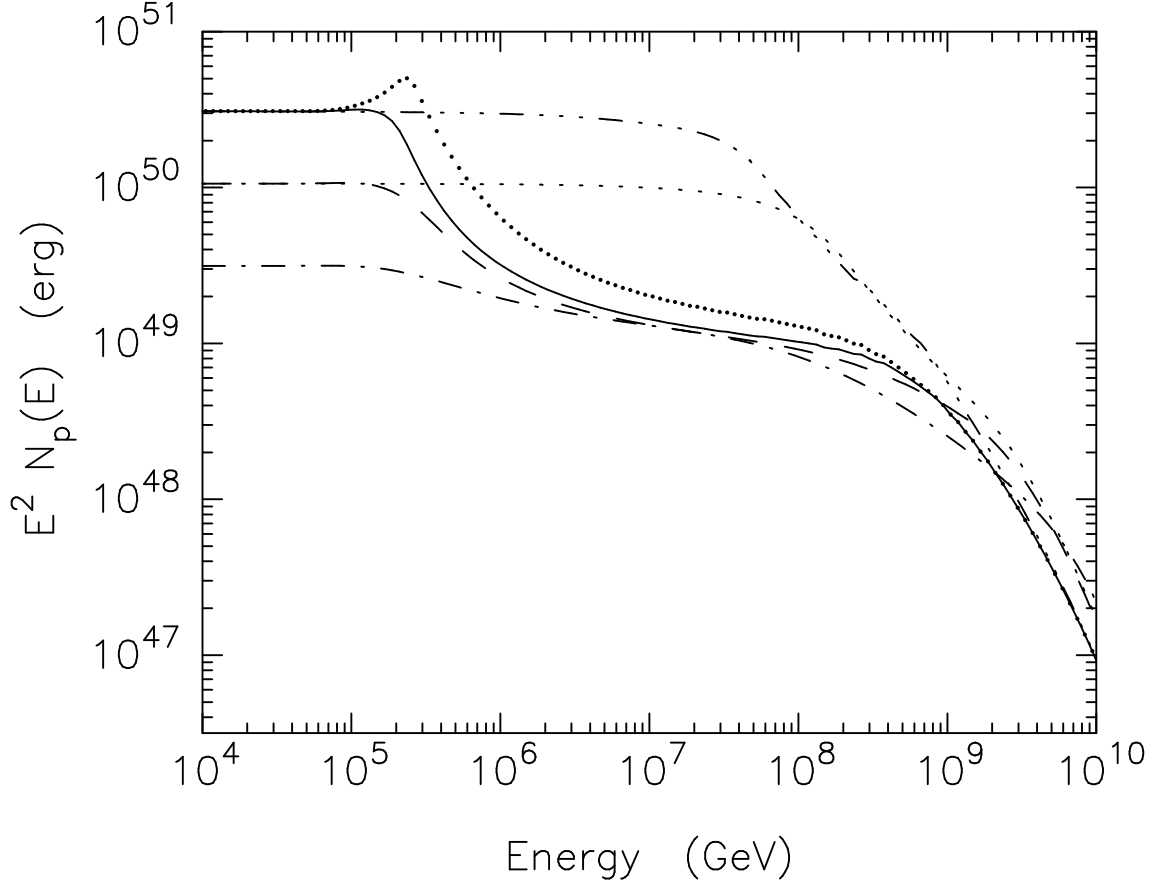}}
\caption{
The energy distribution of relativistic protons, calculated
for the same jet Lorentz-factors $\delta = 7 $, 10, and 15 and
assumptions for the external and internal radiation fields as made in
Fig.\ 1, assuming a power-law injection of relativistic protons with
number index $\alpha_{\rm p}=2$ during $\Delta t = 3 \,$ weeks with a
power $L_{\rm p}=2\times 10^{48} \delta^{-4}$ ergs s$^{-1}$. The full dots show the
proton distribution calculated without the effect of neutrons escaping the 
blob.
}
\label{Fig2}
\end{figure}

\begin{figure}
{\includegraphics{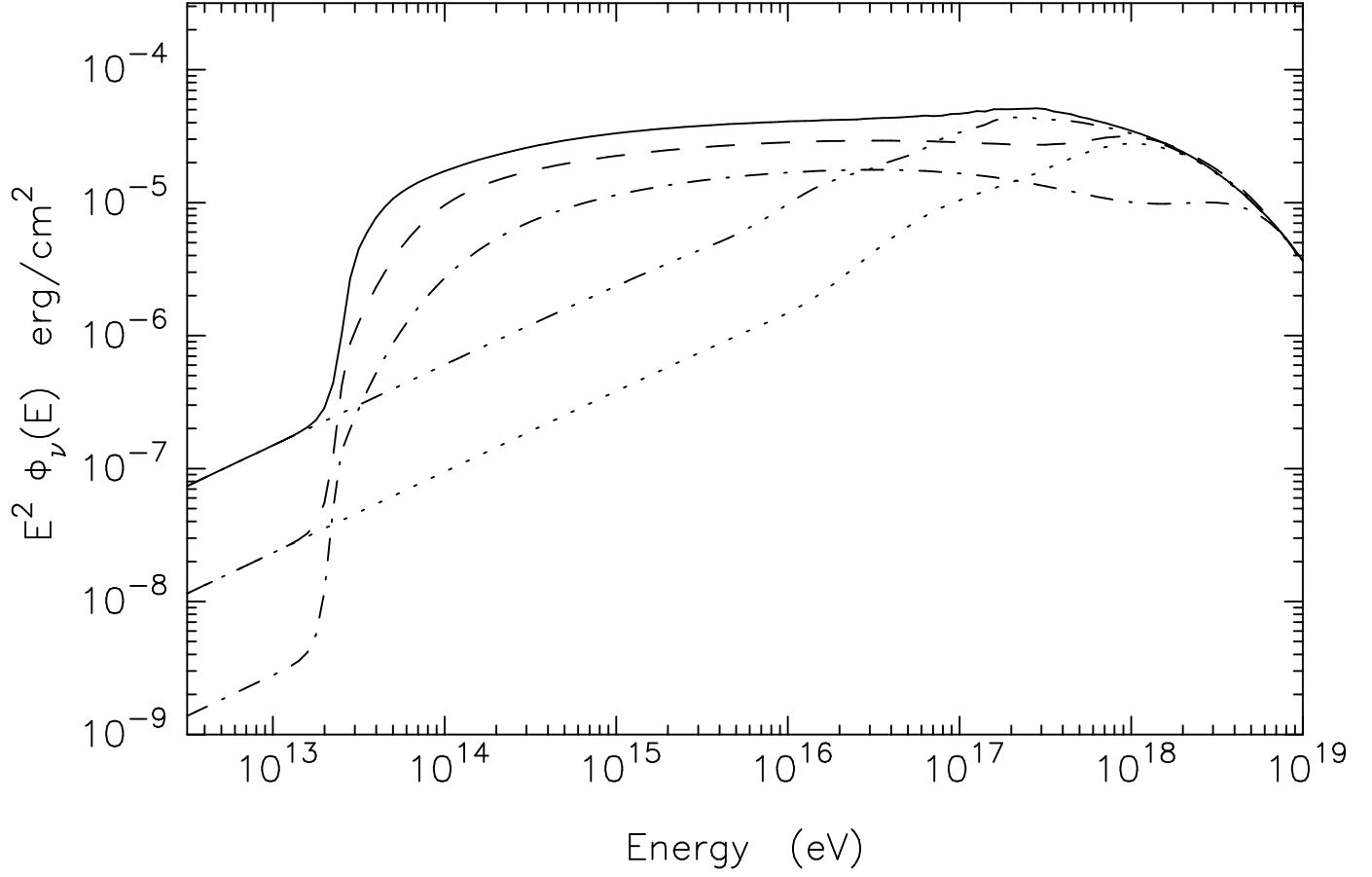}}
\caption{The fluences of neutrinos expected due to photomeson interactions 
of protons shown in Fig.\ 2 with the external (UV) and internal
(synchrotron) photons in the jet of 3C~279 for Dopper factors
$\delta =7$ (solid curve) $\delta = 10 $ (dashed curve), and 
$\delta =15 $ (dot-dashed curve). The dotted and triple-dot--dashed 
curves show the neutrino fluences if the external
photon field in the jet is neglected.}
\label{Fig3}
\end{figure}

\end{document}